\documentclass{ecai} 

\usepackage{latexsym}
\usepackage{amssymb}
\usepackage{amsmath}
\usepackage{amsthm}
\usepackage{booktabs}
\usepackage{enumitem}
\usepackage{graphicx}
\usepackage{color}

\newcommand{\BibTeX}{B\kern-.05em{\sc i\kern-.025em b}\kern-.08em\TeX}

\begin{document}


\begin{frontmatter}


\paperid{123} 


\title{\textit{Value Lens}: Using Large Language Models \\to Understand Human Values}


\author[A,B]{\fnms{Eduardo}~\snm{de la Cruz Fernández}%
  \orcid{0009-0009-2691-4330}%
  \thanks{Corresponding Author. Email: eduardo.cruz@urjc.es}}

\author[B]{\fnms{Marcelo}~\snm{Karanik}\orcid{0000-0001-8848-3681}}

\author[B]{\fnms{Sascha}~\snm{Ossowski}\orcid{0000-0003-2483-9508}}

\address[A]{Universidad Politécnica de Madrid, Madrid, Spain} 
\address[B]{CETINIA, Universidad Rey Juan Carlos, Madrid, Spain}


\begin{abstract}
The autonomous decision-making process, which is increasingly applied to computer systems, requires that the choices made by these systems align with human values. In this context, systems must assess how well their decisions reflect human values. To achieve this, it is essential to identify whether each available action promotes or undermines these values. This article presents \textit{Value Lens}, a text-based model designed to detect human values using generative artificial intelligence, specifically Large Language Models (\textit{LLM}s). The proposed model operates in two stages: the first aims to formulate a formal theory of values, while the second focuses on identifying these values within a given text. In the first stage, an \textit{LLM} generates a description based on the established theory of values, which experts then verify. In the second stage, a pair of \textit{LLM}s is employed: one \textit{LLM} detects the presence of values, and the second acts as a critic and reviewer of the detection process. The results indicate that \textit{Value Lens} performs comparably to, and even exceeds, the effectiveness of other models that apply different methods for similar tasks.
\end{abstract}

\end{frontmatter}


\section{Introduction}

As Artificial Intelligence (AI) systems are increasingly used in daily life to act autonomously, efforts are underway to ensure that their decisions are not only in the interest of the systems' stakeholders, but also respect moral values and ethical considerations. Still, explicitly modelling human values and determining as to how far decisions are aligned with them is a challenging task.

Diverse theories, often rooted in Psychology \citep{Gouveia2014, Maio2010,rokeach1967survey, SCHWARTZ19921}, define human values, their characteristics, and how they interact. Several computational models make proposals as to how to incorporate human values into AI decision-making \citep{SAC2024, Weide2010, Wyner2024, SAATY2010963, Nurwi2021}. To this respect, a key aspect is to determine how a decision or course of action promotes or demotes certain values. For this purpose, a model must first detect which values are relevant in a decision context, to then determine their degree of alignment with regard to the decision alternatives. This problem has been addressed in several approaches \citep{zhu2025,huang2025,fang2023epicurussemeval2023task4,zhang2023maozedongsemeval2023task4,schroter2023adamsmithsemeval2023task4,saha2023rudolfchristopheuckensemeval2023}, some of which were proposed in the Human Value Detection 2023 at SemEval-2023 conference \citep{kiesel-etal-2023-semeval}, whose objective is to use Natural Language Processing techniques to detect human values in text. In parallel, Large Language Models (\textit{LLM}s) have also been explored in creative, value-laden domains \citep{delacruz2024synthetic}.

This article describes \textit{Value Lens}, which relies entirely on \textit{LLM}s to identify human values within a text. The model operates in two stages: the first stage involves conceptualising a specific theory of values, while the second stage focuses on detecting these values. This detection is carried out using a dual approach, which involves using two \textit{LLM}s for detection and confirmation.

The remainder of the article is organized as follows: Section~\ref{model} outlines the \textit{Value Lens} model, Section~\ref{sim&resul} presents the tests conducted and the results achieved, and Section~\ref{conclusions} summarizes the key points of the proposed model.


\section{\textit{Value Lens}} \label{model}
To introduce the \textit{Value Lens}, the Value Detection Model is explained first, followed by the description of the functional interface. 

\subsection{Value detection model} \label{detect}
\begin{figure}[h]
    \centering
    \includegraphics[width=.7\columnwidth]{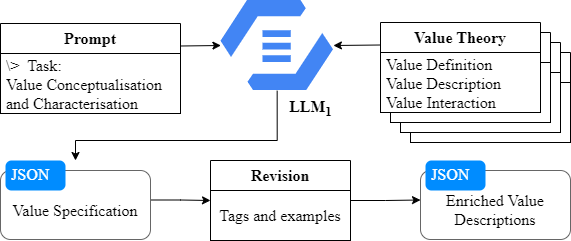}
    \caption{Stage 1: value theory conceptualisation.}
\label{fig:UnderstandingValueTheory}
\end{figure}

The proposed model consists of two main stages: (1) conceptualisation and (2) detecting and analysing human values in text. In the first stage, it learns the key characteristics of a reference value theory, which serves as the foundation for value detection. To enhance the understanding of this value theory, contextual data is used to create a more comprehensive description of values (Fig.~\ref{fig:UnderstandingValueTheory}).

The $\text{\textit{LLM}}_1$ utilises a knowledge transfer prompt to integrate complementary knowledge from various sources. This enables the model to effectively leverage external data and produce more accurate and comprehensive outputs about value specification. The prompt specifies how basic information about each value in the text is extracted. 

Then, each value's name, description, grouping, tags, and examples are stored in a JSON file. Subsequently, a human expert reviews and refines the tags and examples to enhance the quality of the specification. This process also involves adding new tags and examples suggested by the expert. As a result of stage 1, a JSON file with the enriched value specification is returned.

\begin{figure}[h]
    \centering
    \includegraphics[width=.7\columnwidth]{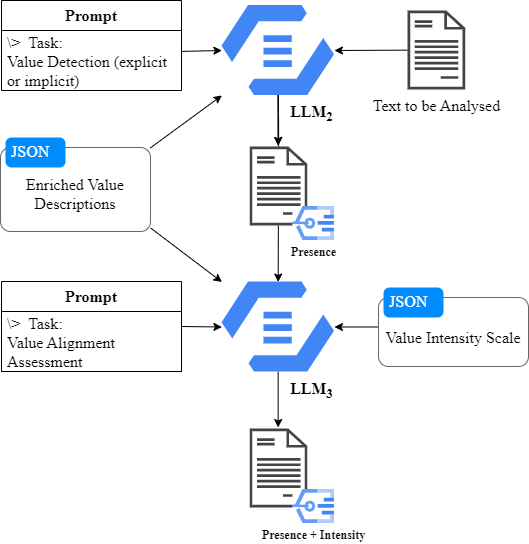}
    \caption{Stage 2: value detection.}
\label{fig:ValueDetectionStage}
\end{figure}

In the second stage (Fig.~\ref{fig:ValueDetectionStage}), a text analysis is conducted based on the value theory conceptualised in the first stage. To accomplish this, a second knowledge transfer prompt is used to specify the task of detecting values by integrating enriched value specifications with the text being analysed. This prompt determines whether the text contains implicit or explicit references to values outlined in value theory.  

The primary goal is to assess how central each identified value is concerning the overall meaning and motivation of the text. As a result, the text is labelled with the list of all values identified by $\text{\textit{LLM}}_2$. While this information is beneficial, it lacks details about the strength of each detected value's relationship with the text. Understanding this intensity is essential to establishing value alignment for effective decision-making processes \citep{khamassi2024strong}. 

A third knowledge transfer prompt is used for the $\text{\textit{LLM}}_3$ to obtain those intensities. The task defined in this prompt integrates the labelled text, returned by the $\text{\textit{LLM}}_2,$ and an intensity scale for values. The $\text{\textit{LLM}}_3$ must determine the degree to which the text promotes or demotes each value. This evaluation should be grounded in textual evidence such as rhetorical emphasis, emotional tone, framing, repetition, placement, and any normative language that affirms or undermines the value. The intensity scale considers the following levels:

\begin{itemize}
    \item \textbf{Strong support}: The text passionately promotes and defends the value, emphasising its importance. This value is central to the message, reinforced with emotional, moral, and logical urgency.
    \item \textbf{Mild support}: The text clearly but gently aligns with the value through positive mention or subtle endorsement, without significant elaboration, insistence, or emphasis.
    \item \textbf{Neutral}: The text mentions the value without signalling any clear support or opposition. The tone is factual, balanced and incidental.
    \item \textbf{Mild resistance}: The text subtly questions, downplays, or introduces alternative perspectives to its value. This opposition is indirect, hedged, or expressed through soft scepticism.
    \item \textbf{Strong resistance}: The text challenges, criticises, or undermines the value directly and forcefully. This includes explicit argumentation, a negative emotional tone, or repeated rejection.
    \item \textbf{Reframing}: The text acknowledges its value but redirects its meaning and context. It introduces a new perspective that shifts emphasis without expressing outright support or opposition.
    \item \textbf{No values}: The text is technical or descriptive, lacking evaluative.
\end{itemize}

At the end of this stage, the text includes labels that indicate the values and intensities associated with it. A justification is also required, providing a clear explanation supported by textual evidence. This approach offers a comprehensive perspective on the insights regarding the values in the text.

\subsection{Functional Interface} \label{interface}

The \textit{Value Lens} interface facilitates the characterisation of a value theory and how its values are referred to in text. For this purpose, several academic papers on a specific value theory are carefully selected for conceptualisation. 
Once the $\text{\textit{LLM}}_1$ generates the specifications based on the documents, the interface displays these specifications in an organised manner for review.

Experts evaluate the conceptual consistency of the results. They make the necessary adjustments if they identify any errors or discrepancies among the concepts, tags, and examples. Additionally, if they find more suitable tags or examples, they incorporate them into the specification to enhance the overall conceptual understanding. This conceptualisation process serves as a preliminary step for one-time value discovery. However, the new document input and expert review must interact continuously. In other words, the aim is to promote continuous improvement in defining the value theory.

On this basis, the text to be analysed is loaded from the interface and, using the enriched value descriptions, it is examined by the $\text{\textit{LLM}}_2$. After analysis, the values detected are used as a label associated with the text. The values in the label do not have a particular order; they are simply a value presence indicator. 

Then, the labelled text and the value intensity scale are used by the $\text{\textit{LLM}}_3$ to determine the alignment of the text concerning the detected values. This description includes values, their intensities, and explanations of the alignments.

The complete \textit{Value Lens} code is available at the following repository: \mbox{\url{https://github.com/segoedu/value-lens}}, and a video describing the system’s functionality can be found at \mbox{\url{https://www.youtube.com/watch?v=TevkbV_W9Ls}}.

\section{Simulation and Results} \label{sim&resul}

This section presents a series of tests to evaluate the model for detecting and analysing human values in text. Then, the results obtained are compared with those of other detection models. Simulations and result analysis have been made in Python3 using the Google Colaboratory environment \citep{googlecolab}. The queries were made using the Groq API interface \citep{groqapi} with the meta-llama/llama-4-scout-17b-16e-instruct language model (with temperature = 0.0).

\subsection{Data Preprocessing and Value Detection} \label{data}

For simulations, Schwartz's value theory \cite{Schwartz1994, SCHWARTZ19921} was selected not only because it is broadly accepted in psychology, but also as a text dataset labelled with Schwartz's value definition is available. It has also been used widely, allowing for measuring and comparing the proposed model's performance.

In the conceptualisation stage, two articles on Schwartz's value theory are utilised: one presents the foundational model of ten basic values \citep{schwartz2012overview}, while the other expands this model to include nineteen values \citep{Schwartz2001}. These two descriptions offer sufficient background to understand value theory. This theory defines each value and presents a model illustrating how values interact based on their motivational compatibilities and incompatibilities. Experts have added tags and examples that align with Schwartz's theory of values to enhance the description provided by the $\text{\textit{LLM}}_1$.

In the value detection stage, the model shown in Fig.~\ref{fig:ValueDetectionStage} has been adapted to use the Touché24-ValueEval dataset \citep{touche24valueval} instead of a single text. This dataset contains 59,662 short example texts labelled with the nineteen values of Schwartz's theory. For the simulations, 7,600 short texts, corresponding to the validation, have been used as input for the $\text{\textit{LLM}}_2$. For the detection process, the labels of these examples were removed and passed to the $\text{\textit{LLM}}_2$, ensuring an unbiased process. In the detection phase, the $\text{\textit{LLM}}_2$ creates a labelled dataset, linking the corresponding values to each sample text.

Using the intensity scale and the enriched value description, explained in section 2.1, the $\text{\textit{LLM}}_3$ analyses the $\text{\textit{LLM}}_2$-labelled dataset to determine the alignment of the values detected in each text. This information is incorporated into each label, generating the dataset labelled with the presence plus intensity of the values. After the detection and analysis, the original Touché24-ValueEval dataset is utilised to calculate precision, recall, macro F1-score, and micro F1-score. 

\subsection{Analysis of Results} \label{analysisRes}

The initial assessment aims to compare, using the micro F1-score (Table~\ref{tab:performance}), the \textit{Value Lens} model's performance against the following similar value detection methods: \textit{BERT}-based \citep{VALE2024}, \textit{Hierocles of Alexandria} \citep{legkas:2024}, \textit{Arthur Schopenhauer} \citep{yunis:2024}, Philo of Alexandria \citep{yeste:2024}, SCaLAR NITK \citep{k:2024} and \textit{Edward Said} \citep{aydin:2024}. 

\begin{table}[h] 
    \caption{Performance Comparison.}
    \centering
    \begin{tabular}{lc}
        \toprule
        \textbf{Model} & \textbf{Micro F1-score} \\
        \midrule
        \textit{Hierocles of Alexandria} & \textbf{0.390} \\
        \textit{\textit{Arthur Schopenhauer}} & 0.350 \\
        \textit{Value Lens} & 0.328 \\
        \textit{BERT}-based & 0.307 \\
        \textit{Philo of Alexandria} & 0.290 \\
        \textit{SCaLAR NITK} & 0.280 \\
        \textit{Edward Said} & 0.280 \\
        \bottomrule
    \end{tabular}
    \label{tab:performance}
\end{table}

Although \textit{Hierocles of Alexandria} obtains the maximum value of F1-score micro, the \textit{Value Lens} model shows adequate behaviour with a value close to the second-best model (\textit{Arthur Schopenhauer}). Table~\ref{tab:performancePerValue} shows the results for each value concerning the best model to give a more precise idea of the proposed model's performance.

\begin{table}[h] 
    \caption{Performance Comparison per Value.}
    \centering
    \begin{tabular}{lcc}
        \toprule
        \textbf{Value} & \textbf{\textit{Hierocles of Alexandria}} & \textbf{\textit{Value Lens}}\\
        \midrule
        Self-direction: thought & 0.150 & \textbf{0.160} \\
        Self-direction: action & \textbf{0.270} & 0.260 \\
        Stimulation & \textbf{0.300} & 0.140 \\
        Hedonism & 0.370 & \textbf{0.400} \\
        Achievement & \textbf{0.450} & 0.440 \\
        Power: dominance & \textbf{0.420} & 0.390 \\
        Power: resources & \textbf{0.490} & 0.280 \\
        Face & \textbf{0.310} & 0.220 \\
        Security: personal & \textbf{0.420} & 0.100 \\
        Security: societal & \textbf{0.490} & 0.370 \\
        Tradition & 0.460 & \textbf{0.540} \\
        Conformity: rules & 0.510 & \textbf{0.530} \\
        Conformity: interpersonal & \textbf{0.240} & 0.140 \\
        Humility & 0.000 & \textbf{0.100} \\
        Benevolence: caring & \textbf{0.340} & 0.290 \\
        Benevolence: dependability & \textbf{0.330} & 0.160 \\
        Universalism: concern & \textbf{0.470} & 0.360 \\
        Universalism: nature & 0.630 & \textbf{0.690} \\
        Universalism: tolerance & \textbf{0.270} & 0.140 \\
        \bottomrule
    \end{tabular}
    \label{tab:performancePerValue}
\end{table}

Although Table 2 shows that the \textit{Hierocles of Alexandria} model obtains better results for detecting most values, the F1-Score of the \textit{Value Lens} model is very close in most classes and even outperforms it in six of them. In addition, \textit{Value Lens} does not have an F1-score at zero, as is the case for the best model for value Humility. 

An interesting aspect is to compare the Value Lens model with these fine-tuned \textit{BERT}-based models that have similar global performance (similar micro F1-score). To do this, one of these models has been selected to compare performance using average precision and recall. \color{black}Table~\ref{tab:MacroMicroAVG} shows that the \textit{BERT}-based model has a precision greater than the recall. This means that when \textit{BERT} predicts a class, it is more likely to be correct, but tends to omit many positive instances. \textit{Value Lens}, on the other hand, tends to be more comprehensive. Its recall is significantly higher than its precision; therefore, it identifies many more positive instances but increases false positives.

\begin{table}[h]
    \caption{Macro and Micro AVG for \textit{Value Lens} and \textit{BERT}-based models.}
    \centering
    \begin{tabular}{lcccc}
        \toprule
        \multicolumn{1}{c}{} & \multicolumn{2}{c}{\textbf{Macro AVG}} & \multicolumn{2}{c}{\textbf{Micro AVG}} \\
        \midrule
        \textbf{Model} & \textbf{Precision} & \textbf{Recall} & \textbf{Precision} & \textbf{Recall}\\
        \midrule
        \textit{Value Lens} & 0.320 & 0.400 & 0.250 & 0.480\\
        \textit{BERT}-based & 0.340 & 0.190 & 0.400 & 0.250\\
        \bottomrule
    \end{tabular}
    \label{tab:MacroMicroAVG}
\end{table}

In addition, the \textit{Value Lens} model has an F1-Score macro of 0.301 against 0.232 of the \textit{BERT}-based model. The macro F1-score averages the F1-score of each class without weighting by the number of instances. A higher macro F1-score suggests that the \textit{Value Lens} model is generally better at balancing accuracy and recall across the different categories, especially for the less frequent classes. This underscores the model's ability to better generalise across different classes, including those with fewer training examples.

\section{Conclusions} \label{conclusions}

This paper describes the \textit{Value Lens} model based on \textit{LLM}s for human value detection and compares its performance to similar models. \textit{Value Lens} offers a promising approach to value alignment analysis that differs from traditional models. 

Unlike classical methods, this model does not require a training stage to identify values in the text. Instead, it incorporates a stage of conceptualisation based on a specific theory of values, enabling the detection of the values and their intensities within a given text. 

The results show that \textit{Value Lens} performs in line with similar models. It also has competitive performance and even exceeds the results of the best model analysed in several categories of values.

Specifically, against fine-tuned models, with similar global performance, \textit{Value Lens} is a superior detection model, as evidenced by its higher macro F1 and better performance in most individual classes.     

Another noteworthy aspect is that \textit{Value Lens} can identify the intensity of each detected value's presence, thanks to its structure of a second \textit{LLM} that criticises the detection results. This improves decision-making by considering value promotion or demotion.



\begin{ack}
Work been supported by grant VAE: TED2021-131295B-C33 funded by MCIN/AEI/10.13039/501100011033 and by the “European Union NextGenerationEU/PRTR”, and by grant COSASS: PID2021-123673OB-C32 funded by MCIN/AEI/10.13039/501100011033 and by “ERDF A way of making Europe”.
\end{ack}



\bibliography{mybibfile}

\end{document}